\documentclass[11pt]{article}

\usepackage[T1]{fontenc}
\usepackage{graphicx}
\usepackage{placeins}
\usepackage{xcolor}
\usepackage{amsmath,amssymb}
\usepackage{geometry}
\usepackage{microtype}
\usepackage{caption}
\usepackage{adjustbox}
\usepackage{hyperref}

\hypersetup{
  colorlinks=true,
  linkcolor=black,
  citecolor=black,
  urlcolor=blue,
  pdftitle={ORION-CMR: On-scanner Reporting with Integrated Foundation Model for End-to-End Cardiac MRI Analysis and Interpretation}
}

\newcommand{\keywords}[1]{%
  \par\vspace{0.6em}\noindent\textbf{Keywords: }%
  \begingroup\def\and{\hspace{0.8em}}#1\endgroup\par
}
\newenvironment{credits}{}{}
\newcommand{\discintname}{Disclosure of Interests.}

\begin{document}

\title{ORION-CMR: On-scanner Reporting with Integrated Foundation Model for End-to-End Cardiac MRI Analysis and Interpretation}

\author{%
\parbox{0.96\textwidth}{\centering
Omer Burak Demirel\textsuperscript{1,2,*},
Kelly K. Horst\textsuperscript{2,*},
Alessio Perazzolo\textsuperscript{2},
Elisa Bruno\textsuperscript{2},
Kenan Kaya\textsuperscript{2},
Rongzhen Ouyang\textsuperscript{2},
Enas Ahmed\textsuperscript{2},
Jouke Smink\textsuperscript{1},
Spencer L. Waddle\textsuperscript{1,2},
Zainudeen Kallumpurath\textsuperscript{1},
Tzu Cheng Chao\textsuperscript{2},
Dinghui Wang\textsuperscript{2},
Steve G. Langer\textsuperscript{2},
Timothy L. Kline\textsuperscript{2},
Panagiotis Korfiatis\textsuperscript{2},
Jacinta Browne\textsuperscript{2},
Ivana Isgum\textsuperscript{2},
Tim Leiner\textsuperscript{2}\\[0.8em]
\small \textsuperscript{1}MR Clinical Science, Philips North America, MN, USA\\
\small \textsuperscript{2}Department of Radiology, Mayo Clinic Rochester, Rochester, MN, USA\\
\small \textsuperscript{*}Equal contribution
}}

\date{}
\maketitle
\vspace{-1.2em}

\begin{abstract}

Cardiovascular magnetic resonance (CMR) provides comprehensive cardiac assessment but remains underutilized because of the complexity of acquisition, post-processing, and interpretation. Existing artificial intelligence (AI) methods address isolated tasks, limiting clinical integration. We present ORION-CMR (On-scanner Reporting with Integrated fOunda-tioN Model), the first clinically evaluated scanner-native end-to-end CMR foundation model. Pretrained on 12,896,733 CMR images from 9,258 studies, ORION-CMR performs sequence classification, ventricular function assessment, late gadolinium enhancement (LGE) detection, binary and multiclass disease classification, and local large language model-based report generation in approximately 90 seconds. The framework was evaluated on public benchmarks and clinically validated in a multi-vendor cohort of 68 subjects with normal examinations, congenital heart disease, dilated cardiomyopathy, and myocardial infarction. ORION-CMR outperformed supervised baselines and the previously published CMR foundation model (CMR-FM), achieving state-of-the-art performance for LGE classification and scar segmentation. Clinical evaluation achieved an AUC of 0.96 for normal-versus-abnormal classification and 0.88 for multiclass disease classification, while generated reports demonstrated 81.4\% agreement with expert interpretation. These results demonstrate the feasibility of real-time scanner-native AI-assisted CMR analysis and automated report generation.


\end{abstract}


%
%
%

\section{Introduction}

Cardiovascular magnetic resonance (CMR) is the reference standard for quantitative assessment of cardiac structure and function, including ventricular volumes and ejection fraction, yet remains underutilized in routine clinical practice \cite{arnold2020cmr,heidenreich2022hf,rajiah2023cmr}. The complexity of CMR acquisition, interpretation, and post-processing, together with the need for specialized expertise and software, limits accessibility compared with other imaging modalities \cite{dweck2016ctcmr}. A recent US national analysis reported fewer than 1,000 physicians interpreting CMR studies, with nearly 70 million Americans living more than 50 miles from a CMR service location \cite{elyaman2025access}. Similarly, increasing demand for CMR has raised concerns about broader adoption worldwide \cite{catapano2024escr,natale2023esr}.

CMR provides comprehensive characterization of cardiac structure, function, and tissue composition without ionizing radiation, enabling longitudinal disease assessment \cite{heidenreich2022hf,bhatt2015congenital,hundley2010consensus}. Late gadolinium enhancement (LGE) imaging identifies myocardial tissue damage with important diagnostic and prognostic value \cite{arnold2020cmr,heidenreich2022hf}. However, comprehensive CMR interpretation requires integrating multiple image sequences, quantitative measurements, and clinical context across fragmented software workflows \cite{dweck2016ctcmr,bhatt2015congenital}, increasing interpretation time, inter-reader variability, and limiting clinical scalability.

Deep learning has substantially improved automated CMR analysis, yet most methods remain designed for a single specific task such as segmentation, classification, or tissue characterization. Foundation models provide unified representations that support multiple downstream tasks from a single encoder \cite{rajiah2023cmr,jacob2025foundation,zhang2025foundation}. This paradigm has improved data efficiency and downstream performance across CT, ultrasound, and MRI \cite{jacob2025foundation,meyer2025ultrasam,hamamci2026ct}. Recent CMR studies have adopted self-supervised Vision Transformers (ViTs) for segmentation and disease classification \cite{zhang2025foundation,jacob2025foundation}. However, these approaches still focus on individual downstream tasks rather than unified, clinically deployable workflows.

In this work, we present ORION-CMR (On-scanner Reporting with Integrated fOundatioN Model), an MRI scanner-native end-to-end CMR interpretation framework. A single pretrained encoder supports sequence classification, ventricular function assessment, LGE classification, binary and multiclass disease classification, and local large language model (LLM)-based report generation. The complete pipeline is implemented directly on the MRI scanner and clinically evaluated, including the generated reports.

\begin{figure}[h]
\centering
\includegraphics[scale=0.5]{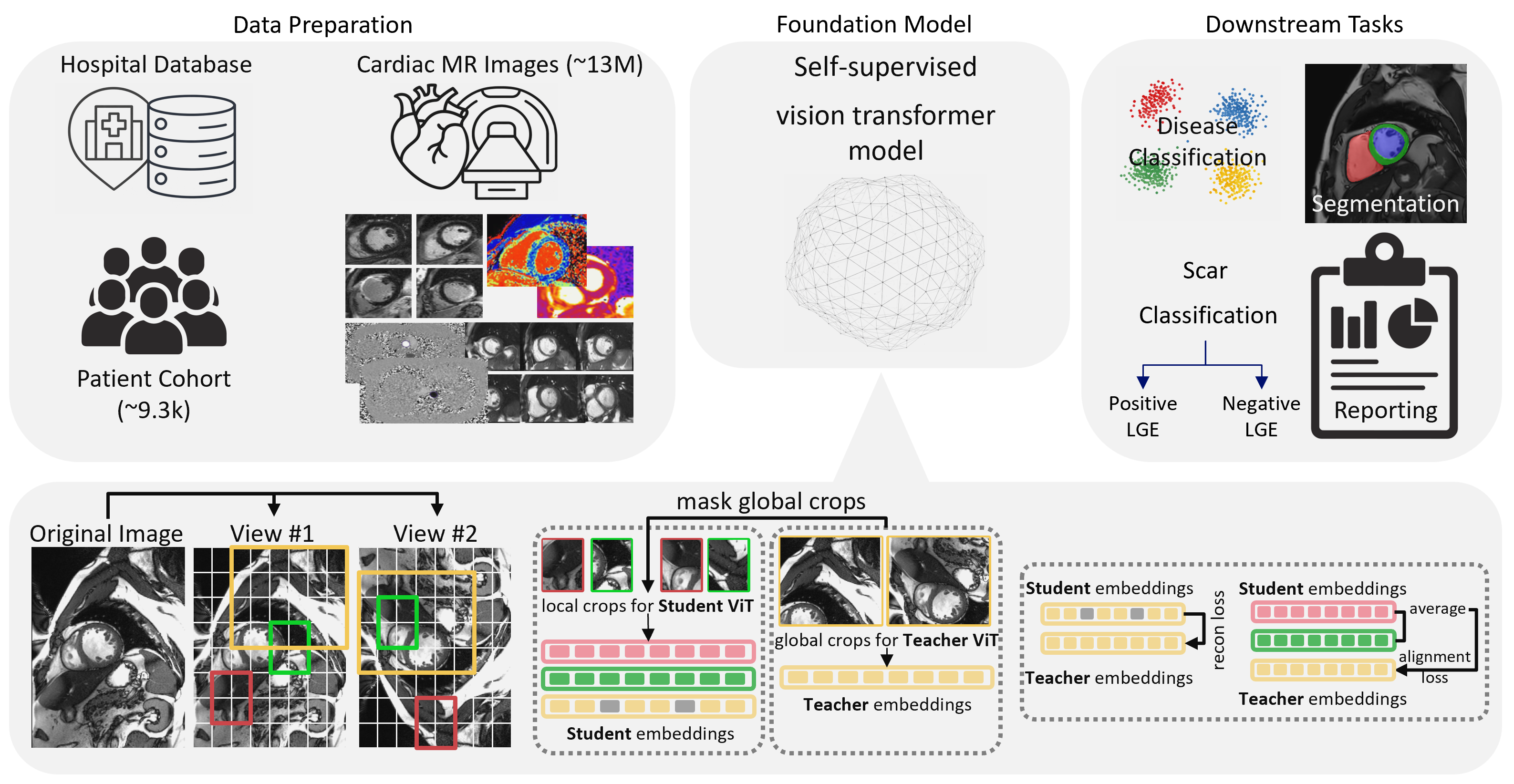}
\caption{Overview of the ORION-CMR (pretrained on 9,258 CMR studies) framework for scanner-native end-to-end CMR analysis and automated report generation.}
\label{fig1}
\end{figure}

\section{Methods}

\subsection{ORION-CMR Framework}

ORION-CMR is a self-supervised CMR foundation model for end-to-end scanner-native CMR interpretation. A shared ViT encoder supports sequence classification, cine short-axis (SAX) segmentation, LGE detection, and disease classification. Quantitative measurements and categorical findings are consolidated into a structured representation that is processed by a locally deployed LLM to generate a clinical report. The complete workflow executes on the MRI scanner in real time. A schematic overview is shown in Fig.~\ref{fig1}.

\subsection{Patient and Data Characteristics}

ORION-CMR was pretrained using 12,896,733 CMR image slices from 9,258 multi-vendor CMR studies spanning localizer, cine, LGE, T1/T2/T2* mapping, perfusion, and phase-contrast flow acquisitions. The mean patient age was 55.5 $\pm$ 17.2 years and 45.9\% (4,247/9,258) of patients were female. The dataset comprised retrospectively collected consecutive CMR studies acquired at Mayo Clinic between 2017 and 2025 on GE Healthcare and Siemens Healthineers scanners.

The Automatic Cardiac Diagnosis Challenge (ACDC) \cite{bernard2018dl} and Evaluation of Myocardial Infarction from Delayed-Enhancement Cardiac MRI (EMIDEC) \cite{lalande2020emidec} datasets were used for public benchmark evaluation. ACDC cine-SAX images were used for disease classification and ventricular segmentation, while EMIDEC phase-sensitive inversion recovery (PSIR)-LGE images were used for infarct classification and scar segmentation. Clinical validation used an in-house multi-vendor cohort of 426 subjects (168 normal, 66 congenital heart disease, 80 dilated cardiomyopathy, and 112 myocardial infarction), partitioned into a 358-subject development cohort for LGE and multiclass disease classification and an independent 68-subject hold-out testing cohort (25 normal, 7 congenital heart disease, 15 dilated cardiomyopathy, and 21 myocardial infarction) for cine-SAX segmentation, LGE classification, binary and multiclass disease classification, and automated report generation.

\subsection{Self-Supervised Foundation Model Training}

Pretraining employed a Self-DIstillation with NO labels (DINO)-style teacher--student framework \cite{caron2021emerging,oquab2024dinov2} with a ViT-Base (ViT-B) backbone and no manual labels. Two variants with patch sizes of $16\times16$ and $8\times8$ pixels (ViT-B16 and ViT-B8, respectively) were investigated to assess whether finer patch representations improve characterization of small anatomical structures and tissue abnormalities in CMR. The student and teacher networks shared the same architecture, with teacher weights updated using an exponential moving average of the student's. 

All CMR images were loaded from a curated study manifest, converted to standardized three-channel inputs, and augmented using a multi-crop strategy with two global and six local views per image. Global views were resized to $224\times224$ pixels, whereas local views were randomly cropped to $96\times96$ pixels. Augmentation comprised percentile-based intensity clipping and normalization, random resized cropping, horizontal flipping, rotation, brightness and contrast jittering, Gaussian blurring, elastic deformation, and additive Gaussian noise. Images were normalized with channel-wise mean and standard deviation values of 0.5. Optimization used AdamW with a base learning rate of $5\times10^{-5}$, cosine learning-rate decay, and a 10-epoch warmup. Teacher momentum increased from 0.996 to 1.0 following a cosine schedule, while teacher temperature increased from 0.04 to 0.07 during warmup. During training, the teacher processed the two global views and the student all eight augmented views. The resulting encoder served as the shared backbone for all downstream tasks.

\subsection{Downstream Tasks}

The pretrained ORION-CMR encoder was transferred to multiple downstream tasks where the foundation model's encoder remained fixed and lightweight task-specific models were trained on top of the learned representations.

\subsubsection{Sequence Classification}

In-house images were categorized into eleven common CMR sequence types by four expert radiologists, including localizer, cine 2-3-4ch-SAX, LGE, flow, perfusion, T1/T2/T2* mapping. The pretrained ViT encoder used concatenated feature embeddings from the final two transformer blocks, yielding a 1536-dimensional representation followed by a multilayer perceptron (MLP) classifier trained using cross-entropy loss and AdamW optimization. Performance was evaluated at the subject level. The predicted sequence labels were subsequently used to automatically identify the required image series for downstream tasks.

\subsubsection{Segmentation}
Cine-SAX and LGE-SAX were performed where input images were first resized to $256\times256$ pixels, intensity-normalized, and converted to three-channel images. Patch tokens from the final four transformer blocks were concatenated, reshaped into spatial feature maps, and processed by a lightweight convolutional decoder to generate full-resolution segmentations. The decoder was trained for 50 epochs using AdamW (learning rate $2\times10^{-4}$, weight decay $10^{-4}$, batch size 8) with a combined class-weighted cross-entropy and soft Dice loss.

For the ACDC benchmark, cine-SAX segmentation targeted the right ventricle (RV), myocardium (Myo), and left ventricle (LV), whereas EMIDEC targeted Myo and scar. As in-house segmentations were unavailable, the ACDC-trained model was applied to the clinical cine-SAX cohort without additional training. The resulting segmentations were used to derive LV and RV parameters, which were incorporated into automated report generation.

\subsubsection{LGE and Disease Classification}

CLS-token embeddings from the pretrained ORION-CMR encoder were extracted, mean-pooled at the subject level, L2-normalized, and classified using calibrated linear support vector machine (SVM) classifiers within a linear-probing framework. Hyperparameters and decision thresholds were optimized exclusively on the validation cohort.

For ACDC, cine-SAX-based disease classification used classification token embeddings from the final four transformer blocks (3072-dimensional). LGE-SAX classification used embeddings from the final two transformer blocks (1536-dimensional) for binary infarct classification for EMIDEC and the in-house cohort. Unlike EMIDEC, which contains only PSIR images, the in-house cohort included PSIR, inversion recovery (IR), and motion-corrected (MOCO) LGE acquisitions, requiring a dedicated LGE-SAX classifier. Because CMR sequence provides complementary anatomical and tissue characterization information, disease classification used cine 2ch, 4ch, SAX, and LGE-SAX examinations. Classification token embeddings from the final two transformer blocks were concatenated into a $4\times1536$-dimensional subject-level representation for multiclass disease classification (normal, congenital heart disease, dilated cardiomyopathy, and myocardial infarction), from which binary normal-versus-abnormal classification was derived.
\subsubsection{Report Generation}
Outputs from ventricular segmentation, LGE classification, and disease classification were consolidated into a structured study representation containing quantitative measurements and AI model predictions. Ventricular volumes, ejection fractions, and indexed left ventricular mass were interpreted using age-, sex-, and ethnicity-specific CMR reference values from the Healthy Hearts Consortium (HHC) 2024 framework to derive standardized clinical findings, including ventricular size, systolic function, and myocardial mass classifications \cite{RaisiEstabragh2024,SCMR2025}. The HHC framework was selected as the primary reference standard because it provides a large contemporary multi-ethnic CMR reference dataset spanning the adult age spectrum, while additional normative CMR reference studies informed pipeline development and verification \cite{Zhan2024}.

The resulting structured findings, together with quantitative ventricular measurements, LGE classifier outputs, and disease classifier predictions, were assembled into a deterministic internal representation of the examination. Automated ventricular measurements were obtained using deep learning–based segmentation, while myocardial tissue characterization was derived from dedicated LGE classifiers \cite{Ruijsink2020,Wang2024}. This verified structured representation was then provided to a locally deployed Qwen2.5-14B-Instruct LLM running through Ollama, which synthesized a complete radiologist-style CMR report from the protected findings \cite{Savage2025}. 

\subsection{Evaluation}

For image analysis tasks, ORION-CMR was compared with supervised baseline models, the previously published CMR cardiac foundation model (CMR-FM) \cite{jacob2025foundation}, and reported state-of-the-art (SoTA) results on the public benchmark datasets when available \cite{bernard2018dl,lalande2022deep}. Supervised baselines consist of a ResNet-50 classifier initialized with ImageNet weights for sequence classification \cite{he2016deep}, U-Net architectures for segmentation \cite{ronneberger2015u}, and ResNet-18 classifiers for disease and LGE classification \cite{he2016deep}. All baseline models were trained and evaluated using identical dataset partitions, preprocessing, and evaluation metrics.

Segmentation performance was assessed using the Dice index (D) and Hausdorff Distance (d$_H$ (mm)). Sequence, LGE, and disease classification performance were evaluated using area under the curve (AUC), accuracy, sensitivity, and specificity. Generated reports were independently evaluated against original reports by three expert readers using a three-category agreement scale: (1) concordant (no or minimal likelihood of changing treatment decisions), (2) intermediate (possible impact on treatment decisions), and (3) discordant (high likelihood of different treatment decisions). Inter-reader agreement was assessed using Cohen’s $\kappa$ for pairwise comparisons. Evaluation focused on the report impression, which contains the key clinical message. Confidence intervals and statistical comparisons were performed using bootstrap resampling and paired significance testing where appropriate ($P<0.05$ was considered statistically significant).

For report evaluation, the LLM-generated reports were compared directly with the original clinical reports authored by expert radiologists. This design enabled assessment of whether an LLM could synthesize verified quantitative measurements and structured findings into clinically coherent reports that closely resembled expert radiologist reports while maintaining factual consistency and minimizing unsupported inferences or hallucinations \cite{Busch2025,Woznicki2025}.

\subsection{On-Scanner Implementation}

The complete ORION-CMR pipeline was deployed on a Philips 1.5T Ambition X MRI scanner equipped with an NVIDIA RTX A6000 GPU \cite{demirel2026prime}. Although the clinical validation cohort comprised retrospectively acquired multi-vendor examinations, the identical scanner-native workflow was applied to all studies. As each image series became available, ORION-CMR first performed sequence classification to identify the acquisition type and automatically trigger the corresponding downstream task (e.g., segmentation or LGE/disease classification). Quantitative measurements and categorical findings were progressively accumulated into a structured representation throughout the examination. Following completion of all required sequences, the locally deployed LLM generated a clinical report, which was exported as a new DICOM series without external network communication, preserving patient privacy.

\begin{table}[h]
\caption{Disease classification performance on ACDC cine-SAX and infarct classification performance on EMIDEC LGE-SAX. Best results are shown in bold.}
\label{tab:classification}
\centering
\scriptsize
\begin{adjustbox}{max width=\textwidth}
\begin{tabular}{lcccccc|c}
\hline
\multicolumn{7}{c|}{
\begin{tabular}{@{}c@{}}
\textbf{ACDC}\\
\textbf{Disease (cine-SAX)}
\end{tabular}}
&
\begin{tabular}{@{}c@{}}
\textbf{EMIDEC}\\
\textbf{LGE (LGE-SAX)}
\end{tabular}
\\
\hline
Method & Overall & NORM & DCM & HCM & RV & MINF & Overall \\
\hline

ResNet-18 \cite{he2016deep}
& 0.480 & 0.300 & 0.600 & 0.200 & 0.600 & 0.700
& 0.667 \\

SoTA \cite{lalande2022deep}
& \textbf{0.960} & \textbf{1.00} & 0.899 & \textbf{1.000} & 0.899 & \textbf{1.000}
& 0.920 \\

CMR-FM \cite{jacob2025foundation}
& 0.700 & 0.500 & 0.400 & 0.800 & 0.800 & \textbf{1.00}
& 0.733 \\

ORION-CMR (ViT-B8)
& 0.820 & 0.900 & \textbf{0.900} & 0.800 & 0.700 & 0.800
& \textbf{0.930} \\

ViT-B16 (ablation)
& 0.740 & 0.600 & 0.500 & 0.900 & \textbf{0.900} & 0.800
& 0.870 \\

\hline
\end{tabular}
\end{adjustbox}
\end{table}

\begin{table}[h]
\caption{ACDC cine-SAX and EMIDEC LGE-SAX segmentation performance.Best results are shown in bold. ( Dice (D), Hausdorff Distance (d$_H$), ViT-B = VB)}
\label{tab:acdc_emidec_segmentation}
\centering
\scriptsize
\begin{adjustbox}{max width=\textwidth}
\begin{tabular}{lcccccc|cccc}
\hline
 & \multicolumn{6}{c|}{ACDC cine-SAX} & \multicolumn{4}{c}{EMIDEC LGE-SAX} \\
Method & LV D & LV d$_H$ & Myo D & Myo d$_H$ & RV D & RV d$_H$ 
& Myo D & Myo d$_H$ & Scar D & Scar d$_H$ \\
\hline
U-Net \cite{ronneberger2015u} 
& 0.932 & 12.853 & 0.833 & 29.549 & 0.837 & 30.285 
& 0.817 & 18.655 & 0.395 & 9.182 \\

SoTA \cite{bernard2018dl} 
& \textbf{0.949} & \textbf{7.150} & \textbf{0.922} & \textbf{8.700} & \textbf{0.910} & \textbf{11.650} 
& 0.879 & 13.010 & 0.712 & 3.120 \\

CMR-FM \cite{jacob2025foundation} 
& 0.933 & N/A & 0.879 & N/A & 0.907 & N/A 
& 0.844 & N/A & N/A & N/A \\

ORION-CMR (VB8)
& 0.942 & 8.527 & 0.878 & 16.417 & 0.899 & 13.606 
& \textbf{0.915} & \textbf{6.110} & \textbf{0.799} & \textbf{3.034} \\

VB16 (ablation) 
& 0.921 & 9.897 & 0.834 & 21.561 & 0.877 & 13.829 
& 0.818 & 9.153 & 0.653 & 7.550 \\
\hline
\end{tabular}
\end{adjustbox}
\end{table}

\begin{table}[!h]
\caption{Subject-level sequence classification accuracy on the clinical cohort.
SAX = short-axis; 2C/3C/4C = two-/three-/four-chamber.}
\label{tab:sequence}
\centering
\scriptsize
\setlength{\tabcolsep}{2.2pt}
\renewcommand{\arraystretch}{0.85}

\begin{adjustbox}{max width=\textwidth}
\begin{tabular}{lcccccccccccc}
\hline
Method & All & T2 & T1 & LGE & T2* & Loc. & Perf. & Flow & SAX & 3C & 2C & 4C \\
\hline
ResNet\cite{he2016deep}
& 0.925 & 0.033 & 0.456 & 0.761 & 0.850 & 0.867
& 0.878 & 0.896 & 0.975 & \textbf{1.000} & \textbf{0.974} & \textbf{0.986} \\

ORION
& \textbf{0.992} & \textbf{0.975} & \textbf{0.991} & \textbf{0.990}
& \textbf{1.000} & \textbf{0.999} & \textbf{0.998} & \textbf{0.998}
& \textbf{0.993} & \textbf{1.000} & 0.973 & 0.979 \\
\hline
\end{tabular}
\end{adjustbox}
\end{table}

\section{Results}

\subsection{Public Benchmark Results}
In all benchmarks, ViT-B8 consistently outperformed ViT-B16 (ablation) and was therefore selected for subsequent experiments. For LGE classification (Table~\ref{tab:classification}), ORION-CMR achieved the highest performance on EMIDEC (0.930), surpassing both the previous SoTA (0.920) and CMR-FM (0.733). On ACDC, ORION-CMR exceeded CMR-FM (0.820 vs. 0.700) and significantly outperformed the supervised baseline (0.480, $P<10^{-2}$).

For segmentation (Table~\ref{tab:acdc_emidec_segmentation}), ORION-CMR achieved D and d$_H$ values comparable to the reported SoTA on ACDC, with differences (D, d$_H$) of LV (0.007, 1.377), myocardium (0.044, 7.717), and RV (0.011, 1.956), respectively. ORION-CMR established new SoTA performance, achieving Dice scores of 0.915 for myocardium and 0.799 for scar On EMIDEC, exceeding the previously reported SoTA (0.879 and 0.712, respectively). Compared with the supervised baseline, segmentation performance was significantly improved ($P<10^{-2}$).

\subsection{Report Evaluation}

\begin{figure}[t]
\centering
\includegraphics[width=\textwidth]{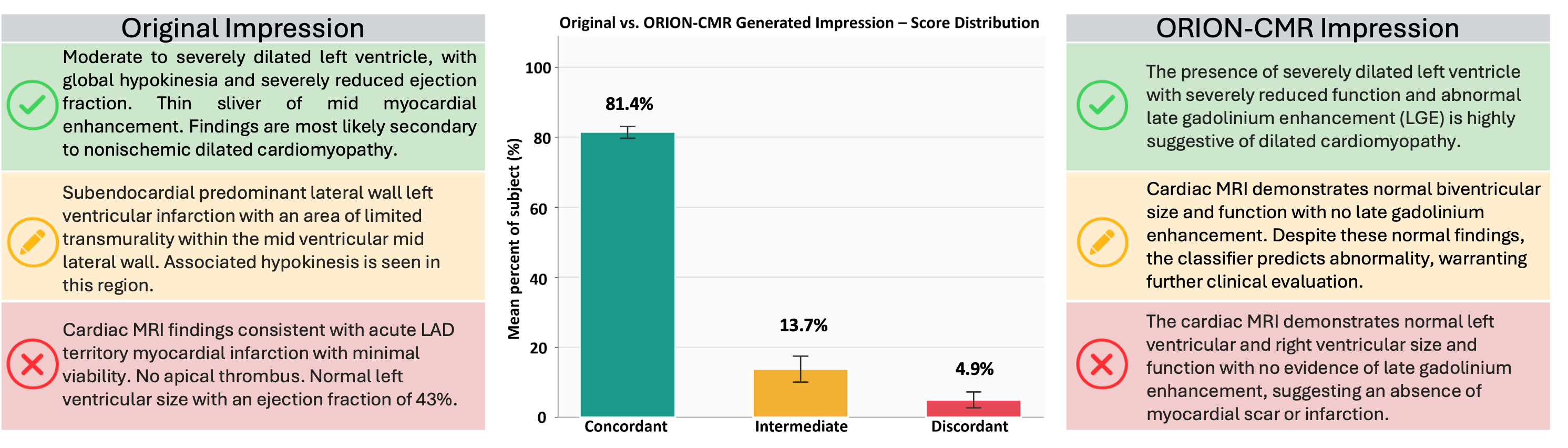}
\caption{Representative examples of original and ORION-CMR generated impressions, together with the distribution of review scores (N=68).}
\label{fig:report_generation}
\end{figure}

\begin{figure}[!h]
\centering
\includegraphics[width=\textwidth]{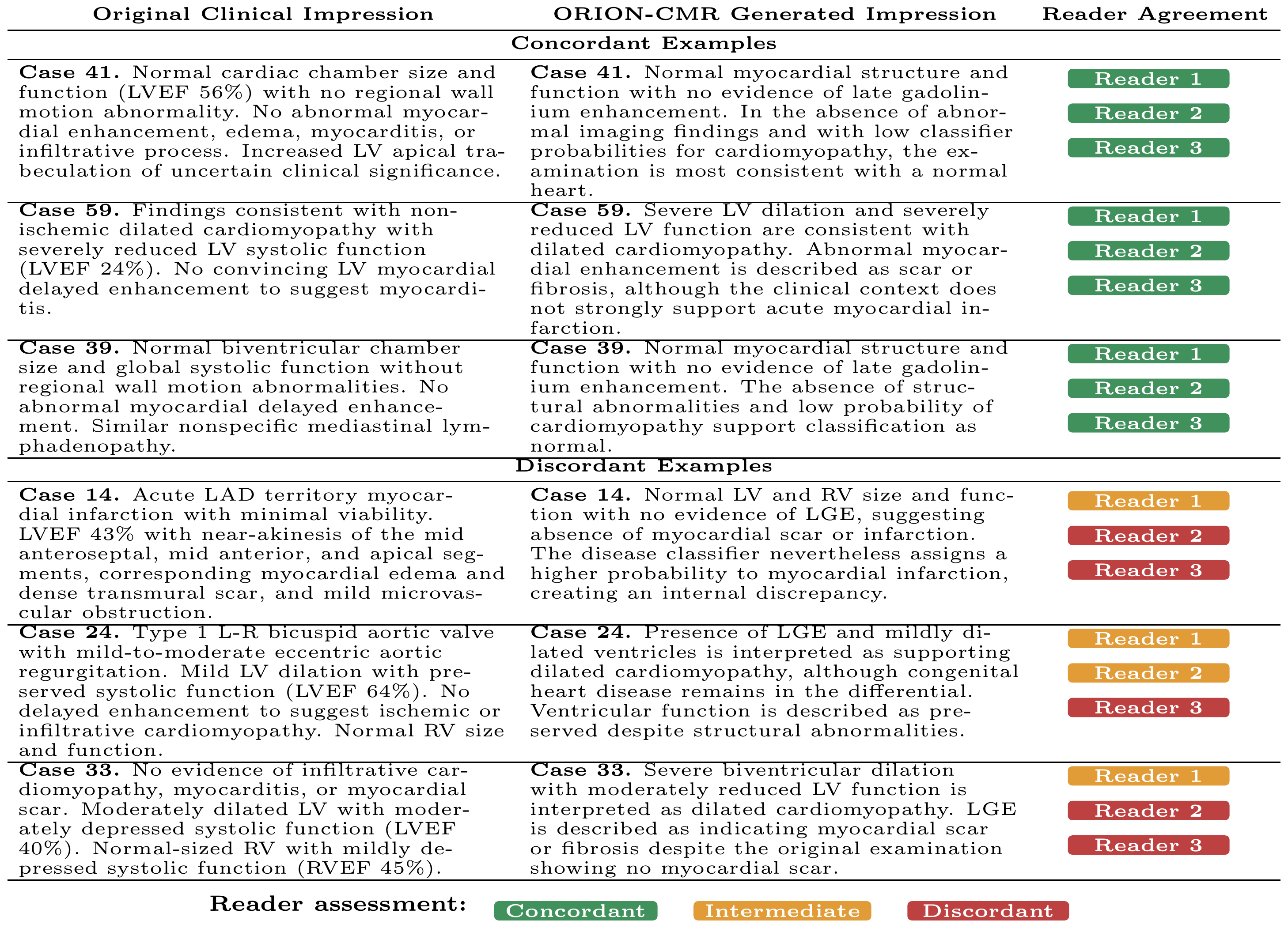}
\caption{Representative original clinical impressions and ORION-CMR-generated impressions with independent expert-reader assessment. Three concordant and three discordant examples are shown. Green, orange, and red indicate concordant, intermediate,
and discordant assessments, respectively.}
\label{fig:report_examples}
\end{figure}

Representative examples of the original and generated reports, together with the radiologist review results, are shown in Fig.~\ref{fig:report_generation}. Among the 68 generated reports, 81.4\% ± 1.7\% were rated as concordant, 13.7\% ± 3.7\% as intermediate, and 4.9\% ± 2.2\% as discordant. Pairwise observed agreement ranged from 76.5\% to 88.2\%, with Cohen's $\kappa$ ranging from 0.41 to 0.61. Representative concordant and discordant cases are shown in Fig.~\ref{fig:report_examples}, comparing the original clinical impression, automatically generated impression, and individual reader assessments. The complete scanner-native ORION-CMR pipeline, from image analysis through automated report generation, required around 90 seconds per subject. The average processing time was $11.2 \pm \mathrm{3.9}$ s for sequence classification, $51.6 \pm \mathrm{15.5}$ s for cine SAX segmentation, $9.7 \pm \mathrm{3.3}$ s for LGE and disease classifications and  $11.2 \pm \mathrm{3.2}$ s for report generation.

\begin{table}[!h]
\caption{Clinical LGE and binary disease classification performance.}
\label{tab:clinical_classification}
\centering
\scriptsize
\begin{adjustbox}{max width=\textwidth}
\begin{tabular}{l|cc|cc}
\hline
& \multicolumn{2}{c|}{\textbf{LGE Classification}} &
\multicolumn{2}{c}{\textbf{Binary Disease Classification}} \\
\hline
Metric &
ResNet-18 \cite{he2016deep} &
ORION-CMR &
ResNet-18 \cite{he2016deep} &
ORION-CMR \\
\hline

AUC
& 0.706 (0.600--0.803)
& \textbf{0.826 (0.722--0.918)}
& 0.761 (0.632--0.874)
& \textbf{0.960 (0.906--0.998)} \\

Accuracy
& 0.650 (0.550--0.740)
& \textbf{0.797 (0.680--0.822)}
& 0.765 (0.667--0.852)
& \textbf{0.912 (0.838--0.971)} \\

Sensitivity
& 0.708 (0.660--0.885)
& \textbf{0.774 (0.622--0.913)}
& 0.804 (0.691--0.906)
& \textbf{0.954 (0.881--1.000)} \\

Specificity
& 0.520 (0.377--0.661)
& \textbf{0.816 (0.686--0.927)}
& 0.680 (0.476--0.864)
& \textbf{0.840 (0.682--0.964)} \\

\hline
\end{tabular}
\end{adjustbox}
\end{table}

\begin{table}[!h]
\caption{Per-class disease classification performance (one-vs-rest AUC).}
\label{tab:disease_perclass}
\centering
\scriptsize
\begin{adjustbox}{max width=\textwidth}
\begin{tabular}{lcc}
\hline
Class & ResNet-18 \cite{he2016deep} & ORION-CMR \\
\hline
Normal
& 0.761 (0.633--0.879)
& \textbf{0.960 (0.906--0.998)} \\

Congenital heart disease
& 0.821 (0.730--0.913)
& \textbf{0.848 (0.667--1.000)} \\

Dilated cardiomyopathy
& 0.674 (0.517--0.814)
& \textbf{0.845 (0.691--0.967)} \\

Myocardial infarction
& 0.720 (0.594--0.836)
& \textbf{0.877 (0.767--0.964)} \\
\hline
\end{tabular}
\end{adjustbox}
\end{table}

\section{Discussion}

This study presents ORION-CMR, a self-supervised CMR foundation model, enabling MRI scanner-native end-to-end CMR interpretation. A single pretrained encoder supports multiple downstream tasks, including sequence classification, segmentation, LGE detection, disease classification, and automated report generation.

Across public benchmarks, ORION-CMR consistently outperformed supervised baselines and the previously published CMR-FM. The ViT-B8 encoder consistently outperformed ViT-B16, suggesting that finer patch representations better capture small anatomical structures and focal myocardial abnormalities. The complete workflow was successfully deployed on an MRI scanner using a locally hosted LLM, demonstrating the feasibility of real-time, privacy-preserving AI-assisted CMR interpretation.

Importantly, ORION-CMR was designed to reduce opportunities for unsupported LLM-generated findings. A shared pretrained ViT backbone performs the image-analysis tasks, with outputs consolidated into a deterministic structured representation. The rule-based LLM generates the final report solely from these predefined quantitative measurements and categorical findings, separating image interpretation from language generation.

This study has several limitations. Clinical validation was limited to a relatively small single-center cohort, particularly the Congenital heart disease subgroup (n=7), warranting further validation in larger, multi-institutional cohorts. Ventricular measurements were validated against values documented in the original clinical reports rather than expert-derived contours, which should be evaluated in future studies. Detailed ablation studies of individual pipeline components and further evaluation of the LLM prompting and structured-data schema are also warranted. The complete pipeline required approximately 90 seconds per subject on the scanner equipped with a GPU, supporting the feasibility of scanner-native deployment. Despite these limitations, to our knowledge, ORION-CMR represents the first fully automated, scanner-native CMR framework integrating image analysis, disease characterization, and report generation into a single end-to-end workflow.

In conclusion, ORION-CMR provides a unified framework for scanner-native image analysis and automated reporting, representing an important step toward clinically deployable end-to-end AI-assisted CMR interpretation.

\begin{credits}

\subsubsection{\discintname}
The authors have no competing interests to declare that are relevant to the content of this article.

\end{credits}

    
    


%
%
%
%
\FloatBarrier

\bibliographystyle{unsrt}
\bibliography{references}





\end{document}